\documentclass[english]{article}
\usepackage[T1]{fontenc}
\usepackage[latin9]{inputenc}
\usepackage{amsbsy}
\usepackage{amstext}
\usepackage{amssymb}
\usepackage{graphicx}
\usepackage{esint}
\usepackage{babel}
\begin{document}
% Full title of the paper (Capitalized)

\title{Topological Quantum Computing and 3-Manifolds}
%\TitleCitation{Topological Quantum Computing and 3-Manifolds}
% Author Orchid ID: enter ID or remove command
%\newcommand{\orcidauthorA}{0000-0003-4813-1010} % Add \orcidA{} behind the author's name
%\newcommand{\orcidauthorB}{0000-0002-7296-7355} % Add \orcidB{} behind the author's name

% Authors, for the paper (add full first names)

\author{Torsten Asselmeyer-Maluga \\ 
German Aerospace Center (DLR), \\ Rosa-Luxemburg-Str. 2, D-10178 Berlin, Germany;\\ torsten.asselmeyer-maluga@dlr.de \\
OrcidId: 0000-0003-4813-1010}

% Contact information of the corresponding author
%\corres{Correspondence: torsten.asselmeyer-maluga@dlr.de}

% Current address and/or shared authorship
%\firstnote{\hangafter=1 \hangindent=1.05em \hspace{-0.82em}\hl{Current address: German Aerospace Center (DLR), Berlin, Germany}} %MDPI: The current address is the same as affiliation 1, please confirm if the current should be removed .
%\secondnote{These authors contributed equally to this work.}
% The commands \thirdnote{} till \eighthnote{} are available for further notes

%\simplesumm{} % Simple summary

%\conference{} % An extended version of a conference paper

% Abstract (Do not insert blank lines, i.e., \\) 
\maketitle
\begin{abstract}
In this paper, we will present some ideas to use 3D topology for quantum
computing. Topological quantum computing in the usual sense works
with an encoding of information as knotted
quantum states of topological phases of matter, thus being locked
into topology to prevent decay. Today, the basic structure is a 2D
system to realize anyons with braiding operations. From the topological
point of view, we have to deal with surface topology. However, usual materials
are 3D objects. Possible topologies for these objects can be more
complex than surfaces. From the topological point of view, Thurston's
geometrization theorem gives the main description of 3-dimensional
manifolds. Here, complements of knots do play a prominent role and
are in principle the main parts to understand 3-manifold topology.
For that purpose, we will construct a quantum system on the complements
of a knot in the 3-sphere. The whole system depends strongly on the
topology of this complement, which is determined by non-contractible,
closed curves. Every curve gives a contribution to the quantum states
by a phase (Berry phase). Therefore, the quantum states can be manipulated
by using the knot group (fundamental group of the knot complement).
The universality of these operations was already showed by M. Planat
et al.
\end{abstract}
%\keyword{topological quantum computing; 3-manifolds; knot complements; braid group}

%\setcounter{section}{-1}

\section{Introduction}

Quantum computing exploits quantum-mechanical phenomena such as superposition
and entanglement to perform operations on data, which in many cases,
are infeasible to do efficiently on classical computers. The basis
of this data is the qubit, which is the quantum analog of the classical
bit. Many of the current implementations of qubits, such as trapped
ions and superconductors, are highly susceptible to noise and decoherence
because they encode information in the particles themselves. Topological
quantum computing seeks to implement a more resilient qubit by utilizing
non-Abelian forms of matter to store quantum information. In such
a scheme, information is encoded not in the quasiparticles themselves,
but in the manner in which they interact and are braided. In topological
quantum computing, qubits are initialized as non-abelian anyons, which
exist as their own antiparticles. Then, operations (what we may think
of as quantum gates) are performed upon these qubits through braiding
the worldlines of the anyons. Because of the non-Abelian nature of
these particles, the manner in which they are exchanged matters (similar
to non-commutativity). Another important property of these braids to
note is that local perturbations and noise will not impact the state
of the system unless these perturbations are large enough to create
new braids. Finally, a measurement is taken by fusing the particles.
Because anyons are their own antiparticles, the fusion will result
in the annihilation of some of the particles, which can be used as
a measurement. We refer to the book in \cite{topoQuantComp} for an introduction
of these ideas.

However, a limiting factor to use topological quantum computing is the
usage of non-abelian anyons. The reason for this is the abelian fundamental
group of a surface. Quantum operations are non-commutative operators
leading to Heisenberg's uncertainty relation, for instance. Non-abelian
groups are at the root of these operators. Therefore, if we use non-abelian
fundamental groups instead of abelian groups, then (maybe) we do not
need non-abelian states to realize quantum computing. {In this paper,
we discuss the usage of (non-abelian) fundamental groups of 3-manifolds for topological quantum computing. At first, quantum gates are elements of the fundamental group represented as $SU(2)$ matrices. The set of possible operations depends strongly on the knot, i.e., the topology. The fundamental group is a topological invariant, thus making this representation of quantum gates part of topological quantum computing. In principle, we have two topological ingredients: the knot and the fundamental group of the knot complement. The main problem now is how these two ingredients can be realized in a quantum system.} 
In contrast to topological quantum computing with anyons,
we cannot directly use {3-manifolds (as submanifolds) like surfaces} in the fractional Quantum Hall effect. Surfaces (or 2-manifolds) embed into a 3-dimensional space like $\mathbb{R}^{3}$ but 3-manifolds require a 5-dimensional
space like $\mathbb{R}^{5}$ as an embedding space. Therefore, we
cannot directly use 3-manifolds. However, as we will argue in the next
section, there is a group-theoretical substitute for a 3-manifolds,
the fundamental group of a knot complement also known as knot group.
Then, we will discuss the knot group of the simplest knot, the trefoil.
The knot group is the braid group of three strands used for anyons
too. The 1-qubit gates are given by the representation of the knot
group into the group $SU(2)$. Here, one can get all 1-qubit operations
by this method. {For an application of these fundamental group representations to topological quantum computing, we need a realization of the
fundamental group in a quantum systems. Here, we will refer to the one-to-one connection between the holonomy of a flat $SU(2)$ connection (i.e., vanishing curvature) and the representation of fundamental group into $SU(2)$. Here, we will use the Berry phase but the corresponding Berry connection admits a non-vanishing curvature. However, we will show that one can rearrange the Berry connection for two-level systems to get a flat $SU(2)$ connection. Then, the holonomy along this connection only depends on the topology of the knot complement so that the manipulation of states are topologically induced. All $SU(2)$-representation of the knot group form the so-called character variety which contains important information about the knot complement (see~in \cite{KirKla:90} for instance). However, the non-triviality of the character variety can be interpreted that knot groups give the 1-qubit operations. 
In Section \ref{sec:Linking-and-2-qubit}, we will discuss the 2-qubit operations by 
linking two knots. Here, every link component carries a representation into 
$SU(2)$. The relation in the fundamental group induces the interaction term.
The interaction terms are known from the Ising model. Therefore, finally we 
get a complete set of operations to realize any quantum circuit: 
a 1-qubit operation by the knot group of the trefoil knot and a 2-qubit 
operation by the complement of the link (Hopf link for instance). } 
For the universality of these operations we refer to the work of 
M. Planat et al. \cite{Planat2018,Planat2019}, which was the main 
inspiration of this work. 
{This paper followed the idea to use knots directly for 
quantum computing. In the focus is the knot complement which is the space 
outside of a knot. Then, the knot is one system and the space outside is a second
system. Currently, the author is working on the concrete realization of this idea.
In this paper, I will present the idea in an abstract manner to clarify whether
knot complements are suitable for quantum computing from informational point of
view.

The usage of knots in physics but also biology is not new. 
One of the pioneers is Louis H. Kauffman, and we refer to his book \cite{KauffmanKnotsPhysics} for many relations between knot theory and natural science. Furthermore, note his ideas about topological information \cite{KauffmanKnotLogic} (see also in \cite{KnotsKauffmanBoi,TopoKnotModelsBoi}). 
Knots are also important models in quantum gravity, see, for instance, in \cite{BaezHighDimAlgebra}, and in particle physics~\cite{BTMarkSmolin2007,Gresnigt2018,AsselmeyerMaluga2019}.}

\section{Some Preliminaries and Motivation: 3-Manifolds and Knot Complements}

The central concept for the following paper is the concept of a smooth
manifold. To present this work as self-contained as possible, we will
discuss some results in the theory of 2- and 3-dimensional manifolds
which is the main motivation for this paper. At first we will give
the formal definition of a manifold:
\begin{itemize}
\item Let $M$ be a Hausdorff topological space covered by a (countable)
family of open sets, ${\mathcal{U}}$, together with homeomorphisms,
$\phi_{U}:U\in{\mathcal{U}}\rightarrow U_{R},$ where $U_{R}$ is
an open set of ${\mathbb{R}}^{n}.$ This defines $M$ as a topological
manifold. For smoothness we require that, where defined, $\phi_{U}\cdot\phi_{V}^{-1}$
is smooth in ${\mathbb{R}}^{n},$ in the standard multivariable calculus
sense. The family ${\mathcal{A}}=\{{\mathcal{U}},\phi_{U}\}$ is called
an atlas or a differentiable structure. Obviously, ${\mathcal{A}}$
is not unique. Two atlases are said to be compatible if their union
is also an atlas. From this comes the notion of a maximal atlas. Finally,
the pair $(M,{\mathcal{A}})$, with ${\mathcal{A}}$ maximal, defines
a smooth manifold of dimension $n$. 
\item An important extension of this construction yields the notion of smooth
manifold with boundary, $M$, defined as above, but with the atlas
such that the range of the coordinate maps, $U_{R},$ may be open
in the half space, ${\mathbb{R}}_{+}^{n}$, that is, the subspace
of ${\mathbb{R}}^{n}$ for which one of the coordinates is non-positive,
say $x^{n}\le0.$ As a subspace of ${\mathbb{R}}^{n},\ {\mathbb{R}}_{+}^{n}$
has a topologically defined boundary, namely, the set of points for
which $x^{n}=0.$ Use this to define the (smooth) boundary of $M,\ \partial M,$
as the inverse image of these coordinate boundary points.
\end{itemize}

In the following, we will concentrate on the special theory of 2-
and 3-manifolds (i.e., manifolds of dimension 2, surfaces, or 3). The
classification of 2-manifolds has been known since the 19th century. In
contrast, the corresponding theory for 3-manifolds based on ideas
of Thurston around 1980 but was completed 10 years ago. In both cases---2- and 3-manifolds---the manifold is decomposed by the operation $M\#N$,
the connected sum.

Let $M,N$ be two $n$-manifolds with boundaries $\partial M,\partial N$.
The connected sum %MDPI: is the italic necessary?
$M\#N$ is the procedure of cutting out a
disk $D^{n}$ from the interior $int(M)\setminus D^{n}$ and $int(N)\setminus D^{n}$
with the boundaries $S^{n-1}\sqcup\partial M$ and $S^{n-1}\sqcup\partial N$,
respectively, and gluing them together along the common boundary component
$S^{n-1}$. 

For 2-manifolds, the basic elements are the 2-sphere $S^{2}$, the
torus $T^{2}$ or the Klein bottle $\mathbb{R}P^{2}$. Then, one gets
for the classification of 2-manifolds:
\begin{itemize}
\item Every compact, closed, oriented 2-manifold is homeomorphic to either
$S^{2}$ or the connected sum 
\[
\underbrace{T^{2}\#T^{2}\#\ldots\#T^{2}}_{g}
\]
of $T^{2}$ for a fixed genus $g$. Every compact, closed, non-oriented
2-manifold is homeomorphic to the connected sum 
\[
\underbrace{{\mathbb{R}}P^{2}\#{\mathbb{R}}P^{2}\#\ldots\#{\mathbb{R}}P^{2}}_{g}
\]
of ${\mathbb{R}}P^{2}$ for a fixed genus $g$.
\item Every compact 2-manifold with boundary can be obtained from one of
these cases by cutting out the specific number of disks $D^{2}$ from
one of the connected sums. 
\end{itemize}

A connected 3-manifold $N$ is prime if it cannot be obtained as a
connected sum of two manifolds $N_{1}\#N_{2}$ neither of which is
the 3-sphere $S^{3}$ (or, equivalently, neither of which is the homeomorphic
to $N$). Examples are the 3-torus $T^{3}$ and $S^{1}\times S^{2}$,
but also the Poincare sphere. According to the work in \cite{Mil:62}, any compact,
oriented 3-manifold is the connected sum of an unique (up to homeomorphism)
collection of prime 3-manifolds (prime decomposition). A subset of
prime manifolds are the irreducible 3-manifolds. A connected 3-manifold
is irreducible if every differentiable submanifold $S$ homeomorphic
to a sphere $S^{2}$ bounds a subset $D$ (i.e., $\partial D=S$) which
is homeomorphic to the closed ball $D^{3}$. The only prime but reducible
3-manifold is $S^{1}\times S^{2}$. 

For the geometric properties (to meet Thurston's geometrization theorem)
we need a finer decomposition induced by incompressible tori. A properly
embedded connected surface $S\subset N$ is called 2-sided (The ``sides''{}
of $S$ then correspond to the components of the complement of $S$
in a tubular neighborhood $S\times[0,1]\subset N$.) if its normal
bundle is trivial, and 1-sided if its normal bundle is nontrivial.
A 2-sided connected surface $S$ other than $S^{2}$ or $D^{2}$ is
called incompressible if for each disk $D\subset N$ with $D\cap S=\partial D$
there is a disk $D'\subset S$ with $\partial D\prime=\partial D$.
The boundary of a 3-manifold is an incompressible surface. Most importantly,
the 3-sphere $S^{3}$, $S^{2}\times S^{1}$ and the 3-manifolds $S^{3}/\Gamma$
with $\Gamma\subset SO(4)$ a finite subgroup do not contain incompressible
surfaces. The class of 3-manifolds $S^{3}/\Gamma$ (the spherical
3-manifolds) include cases like the Poincare sphere ($\Gamma=I^{*}$
the binary icosaeder group) or lens spaces ($\Gamma=\mathbb{Z}_{p}$
the cyclic group). Let $K_{i}$ be irreducible 3-manifolds containing
incompressible surfaces then we can $N$ split into pieces (along
embedded $S^{2}$)
\begin{equation}
N=K_{1}\#\cdots\#K_{n_{1}}\#_{n_{2}}S^{1}\times S^{2}\#_{n_{3}}S^{3}/\Gamma\,,\label{eq:prime-decomposition}
\end{equation}
where $\#_{n}$ denotes the $n$-fold connected sum and $\Gamma\subset SO(4)$
is a finite subgroup. The decomposition of $N$ is unique up to the
order of the factors. The irreducible 3-manifolds $K_{1},\ldots,\,K_{n_{1}}$
are able to contain incompressible tori and one can split $K_{i}$
along the tori into simpler pieces $K=H\cup_{T^{2}}G$ \cite{JacSha:79}
(called the JSJ decomposition). The two classes $G$ and $H$ are
the graph manifold $G$ and hyperbolic 3-manifold $H$.

In 1982, W.P. Thurston presented a program intended to classify smooth
3-manifolds and solve the Poincare conjecture by investigating the
possible geometries on such 3-manifolds. For a survey of this topic
see in \cite{Thu:97}. The key ingredient of this classification ansatz
is the concept of a model geometry. Again, in this section, all manifolds
are assumed to be smooth. 

A model geometry $(G,X)$ consists of a simply connected manifold
$X$ together with a Lie group $G$ of diffeomorphisms acting transitively
on $X$ fulfilling certain set of conditions. One of these is that
there is a $G$-invariant Riemannian metric. For example, reducing
the dimension, we can consider 2-dimensional model geometries of a
2-manifold $X$. From Riemannian geometry, we know that any $G$-invariant
Riemannian metric on $X$ has constant Gaussian curvature (recall
that $G$ must be transitive). A constant scaling of the metric allows
us to normalize the curvature to be $0$, $1$, or $-1$ corresponding
to the Euclidean (${\mathbb{E}}^{2}$), spherical ($S^{2}$) and hyperbolic
(${\mathbb{H}}^{2}$) space, respectively. Thus, there are precisely
three two-dimensional model geometries: spherical, Euclidean, and hyperbolic. 

It is a surprising fact that there are also a finite number of three-dimensional
model geometries. It turns out that there are eight geometries: spherical,
Euclidean, hyperbolic, mixed spherical-Euclidian, mixed hyperbolic-Euclidian,
and three exceptional cases. A geometric structure on a more general manifold
$M$ (not necessarily simply connected) is defined by a model geometry
$(G,X)$ where $X$ is the universal covering space to $M$, i.e., $M=X/\pi_{1}(M)$.
This is equivalent to a representation $\pi_{1}(M)\to G$ of the fundamental
group into $G$. Of course a geometric structure on a 3-manifold may
not be unique but Thurston explored decompositions into pieces each
of which admit a unique geometric structure. This decomposition proceeds
by splitting $M$ into essentially unique pieces using embedded 2-spheres
and 2-tori in such a way that a model geometry can be defined on each
piece. Thus, 
\begin{itemize}
 \item {{Thurston's Geometrization conjecture can be stated:}}%MDPI: is the boldnecessary? and wo format it as list, please confirm.

The interior of every compact 3-manifold has a canonical decomposition
into pieces (described above), which have one of the eight geometric
structures.

In short, every 3-manifold can be uniquely decomposed (long 2-spheres)
into prime manifolds where some of these prime manifolds can be further
split (along 2-tori) into graph $G$ and hyperbolic manifolds $H$.
Then, $G,H$ have a disjoint union of 2-tori as boundary; but how can
we construct these manifolds having a geometric structure? A knot
in mathematics is the embedding of a circle into the 3-sphere $S^{3}$
(or in $\mathbb{R}^{3}$), i.e., a closed knotted curve. Let $K$ be
a prime knot (a knot not decomposable by a sum of two knots). With
$K\times D^{2}$ we denote a thicken knot, {i.e.} %Should it be changed to i.e.,?
a closed knotted solid
torus. The knot complement $C(K)=S^{3}\setminus\left(K\times D^{2}\right)$
is a 3-manifold with boundary $\partial C(K)=T^{2}$. It was shown
that prime knots are divided into two classes: hyperbolic knots ($C(K)$
admits a hyperbolic structure) and non-hyperbolic knots ($C(K)$ admits
one of the other seven geometric structures). An embedding of disjoints
circles into $S^{3}$ is called a link $L.$ Then, $C(L)$ is the link
complement. Here, the situation is more complicated: $C(L)$ can admit
a geometric structure or it can be decomposed into pieces with a geometric
structure. $C(L),C(K)$ are one of the main models for $G$ or $H$
for suitable knots and links. If we speak about 3-manifolds then we
have to consider $C(K)$ as one of the basic pieces. Furthermore,
there is the Gordon--Luecke theorem: if two knot complements are homeomorphic,
then the knots are equivalent (see in \cite{GordonLuecke1989} for the
statement of the exact theorem). Interestingly, knot complements of
prime knots are determined by its fundamental group. For the fundamental
group, one considers closed curves which are not contractible. Furthermore,
two curves are equivalent if one can deform them into each other (homotopy
relation). The concatenation of curves can be made into a group operation
up to deformation equivalence (i.e., homotopy). Formally, it is the
set of homotopy classes $[S^{1},X]$ of maps $S^{1}\to X$ (the closed
curves) into a space $X$ up to homopy, denoted by $\pi_{1}(X)$.
The fundamental group $\pi_{1}(C(K))$ of the knot complement is also
known as knot group. Here, we refer to the books~in \cite{Rol:76,KauffmanKnotsPhysics,PrasSoss:97}
for a good introduction into this theory. The main idea of this paper
is the usage of the knot group as substitute for a 3-manifold and
try to use this group for quantum computing.
 \end{itemize}
 
\section{Knot Complement of the Trefoil Knot and the Braid Group \boldmath{$B_{3}$}}

Any knot can be represented by a projection on the plane with no multiple
points which are more than double. As an example let us consider the
simplest knot, the trefoil knot $3_{1}$ (knot with three crossings).

\begin{figure}
%\widefigure
\includegraphics[scale=0.15]{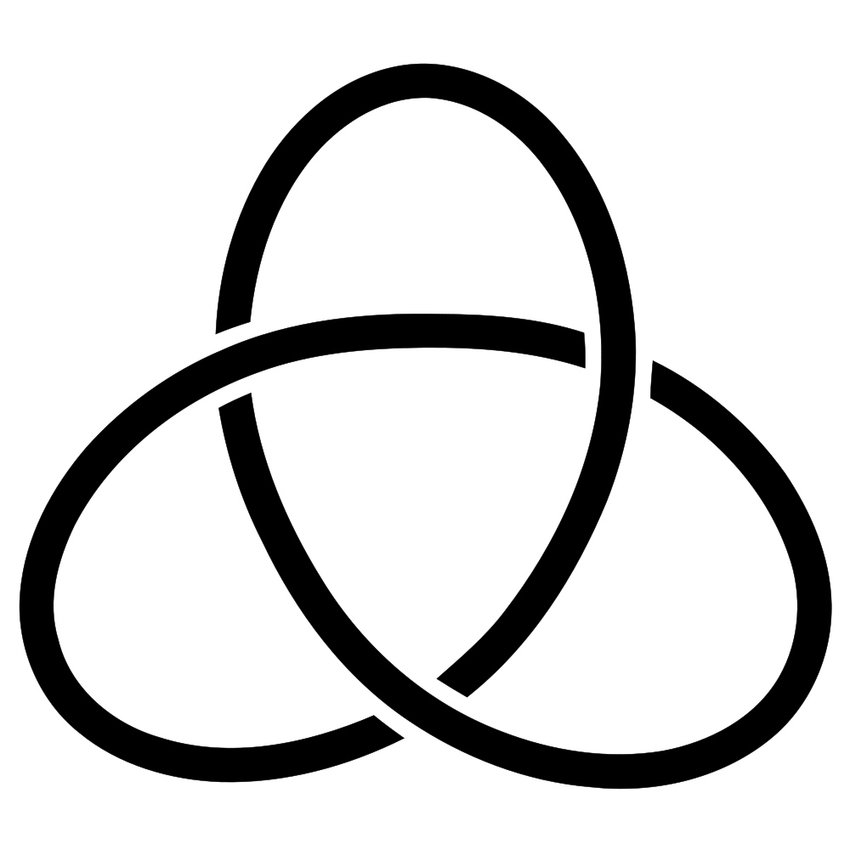}

\caption{The simplest knot, trefoil knot $3_{1}$. \label{fig:trefoil-knot}}

\end{figure}

\begin{figure}
%\widefigure
\includegraphics[scale=0.25]{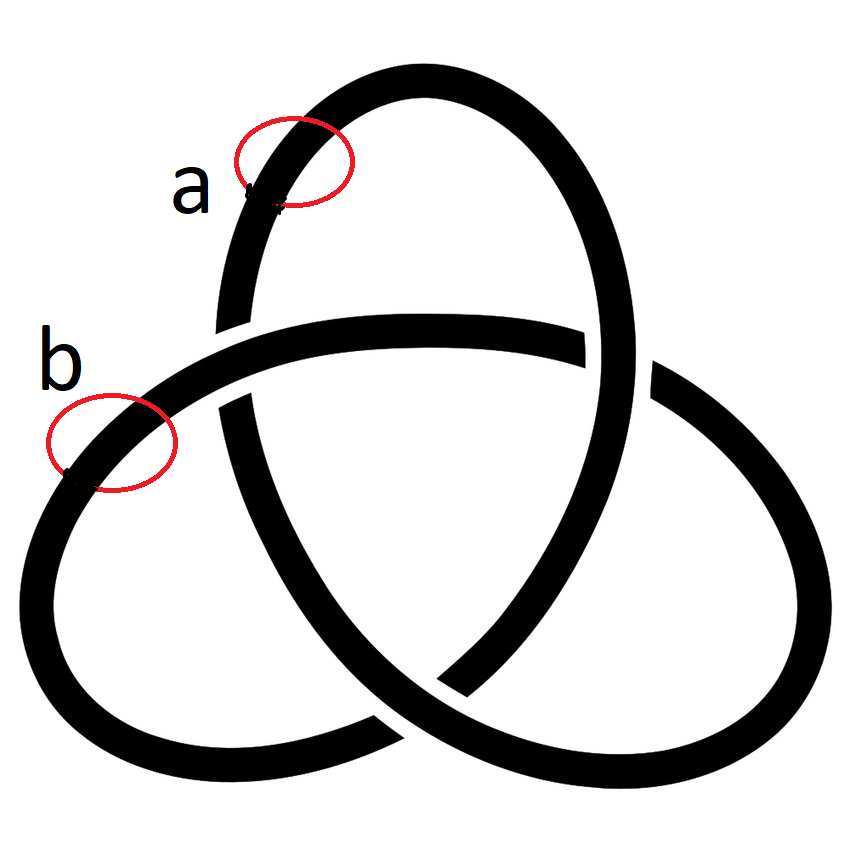}\caption{{Generators (red circle)} $a,b$ of knot group for trefoil $3_{1}$. \label{fig:generator-trefoil}} %mdpi: Need to explain a, b? 
\end{figure}

The plane projection of the trefoil is shown in Figure \ref{fig:trefoil-knot}.
This projection can be divided into three arcs, around each arc we
have a closed curve as generator of $\pi_{1}(C(3_{1}))$ denoted by
$a,b,c$ {(see also Figure \ref{fig:generator-trefoil} for the definition of the generators $a,b$)}. Now each crossing gives a relation between the corresponding
generators: $c=a^{-1}ba,b=c^{-1}ac,a=b^{-1}cb$, i.e., we get the knot
group
\[
\pi_{1}(C(3_{1}))=\langle a,b,c|\,c=a^{-1}ba,b=c^{-1}ac,a=b^{-1}cb\rangle
\]

Then, we substitute the expression $c=a^{-1}ba$ into the other relations
to get a representation of the knot with two generators and one relation.
From relation $a=b^{-1}cb$ we will obtain $a=b^{-1}(a^{-1}ba)b$
or $bab=aba$ and the other relation $b=c^{-1}ac$ gives nothing new.
Finally, we will get the well-known result
\[
\pi_{1}(C(3_{1}))=\langle a,b|\,bab=aba\rangle
\]

However, this group is also well known; it is the braid group $B_{3}$
of three strands. In general, the braid group $B_{n}$ is generated
by $\left\{ 1,\sigma_{1},\ldots,\sigma_{n-1}\right\} $ subject to
the relations (see in \cite{PrasSoss:97})
\[
\sigma_{i}\sigma_{i+1}\sigma_{i}=\sigma_{i+1}\sigma_{i}\sigma_{i+1}\qquad\sigma_{i}\sigma_{j}=\sigma_{j}\sigma_{i}\quad|i-j|>1
\]

For $B_{3}$ we have two generators $\sigma_{1},\sigma_{2}$ with
one relation $\sigma_{1}\sigma_{2}\sigma_{1}=\sigma_{2}\sigma_{1}\sigma_{2}$
agreeing with $\pi_{1}(C(3_{1}))$. The braid group $B_{3}$ has connections
to different areas. Notable is the relation to the modular group $SL(2,\mathbb{Z})$
(the group of integer $2\times2$ matrices with unit determinant)
generated by
\[
S=\left(\begin{array}{cc}
0 & 1\\
-1 & 0
\end{array}\right)\qquad U=\left(\begin{array}{cc}
1 & 1\\
-1 & 0
\end{array}\right)
\]

It is well-known that $B_{3}$ maps surjectively onto $SL(2,\mathbb{Z})$
via the map
\[
\sigma_{1}\mapsto\left(\begin{array}{cc}
1 & 1\\
0 & 1
\end{array}\right)\qquad\sigma_{2}\mapsto\left(\begin{array}{cc}
1 & 0\\
-1 & 1
\end{array}\right)
\]

However, then $\sigma_{1}\sigma_{2}\sigma_{1}=\sigma_{2}\sigma_{1}\sigma_{2}$
maps to $S$ and $\sigma_{1}\sigma_{2}$ maps to $U$. It is interesting
to note that $(\sigma_{1}\sigma_{2})^{3}$ is in the center $Z$ of
$B_{3}$ (i.e., this elements commutes with all other elements) and
$B_{3}/Z=SL(2,\mathbb{Z})/\left\{ \pm1\right\} $, or $B_{3}$ is
the central extension of $PSL(2,\mathbb{Z})=SL(2,\mathbb{Z})/\left\{ \pm1\right\} $
(the $(2,3,\infty)$ triangle group) by the integers $\mathbb{Z}$
(see {\cite{AsselmeyerMaluga2019}} for an application of this relation
in physics).%MDPI: Please confirm which reference this is 
\vspace{-12pt}
\section{Using the Trefoil Knot Complement for Quantum Computing}

In this section, we will get in touch with quantum computing. The main
idea is the interpretation of the braid group $B_{3}$ as operations
(gates) on qubits. From the mathematical point of view, we have to
consider the representation of $B_{3}$ into $SU(2)$, i.e., a homomorphism
\[
\phi:B_{3}\to SU(2)
\]
mapping sequences of generators (called words) into matrices as elements
of $SU(2)$. For completeness we will study some representations. {Here, we follow the work in \cite{KauffmanLomonaco2008} to illustrate the general theory.}
At first, we note that a matrix in $SU(2)$ has the form
\[
M=\left(\begin{array}{cc}
z & w\\
-\bar{w} & \bar{z}
\end{array}\right)\qquad|z|^{2}+|w|^{2}=1
\]
where $z$ and $w$ are complex numbers. Now we choose a well-known
basis of $SU(2)$:
\begin{equation}
\boldsymbol{1}=\left(\begin{array}{cc}
1 & 0\\
0 & 1
\end{array}\right)\quad\boldsymbol{i}=\left(\begin{array}{cc}
i & 0\\
0 & -i
\end{array}\right)\quad\boldsymbol{j}=\left(\begin{array}{cc}
0 & 1\\
-1 & 0
\end{array}\right)\quad\boldsymbol{k}=\left(\begin{array}{cc}
0 & i\\
i & 0
\end{array}\right)
\label{eq:matrix-basis-quaionions}
\end{equation}
so that 
\[
M=a\boldsymbol{1}+b\boldsymbol{i}+c\boldsymbol{j}+d\boldsymbol{k}
\]
with $a^{2}+b^{2}+c^{2}+d^{2}=1$ (and $z=a+bi,w=c+di$). The algebra
of $\boldsymbol{1},\boldsymbol{i},\boldsymbol{j},\boldsymbol{k}$ are known as
quaternions {(with relations $ij=k$, $I^2=j^2=k^2=-1$).
In the following, we will switch between the usual basis $1,i,j,k$ of the
quaternions and the matrix representation with basis 
$\boldsymbol{1},\boldsymbol{i},\boldsymbol{j},\boldsymbol{k}$, 
see (\ref{eq:matrix-basis-quaionions}).} 
Then, the unit quaternions (of length 1) can be identified
with the elements of $SU(2)$. Pure quaternions are defined by all
expressions $b\boldsymbol{i}+c\boldsymbol{j}+d\boldsymbol{k}$ (i.e.,
$a=0$). Now, the homomorphism $\phi$ above is the mapping 
\[
g=\phi(\sigma_{1}),\quad h=\phi(\sigma_{2})
\]
so that $ghg=hgh$. Let $u,v$ be pure quaternions of length $1$
(unit, pure quaternions). Now, for $g=a+bu$ and $h=c+dv$ we have
to choose
\[
g=a+bu,\:h=a+bv,\:bu\cdot bv=a^{2}-\frac{1}{2}
\]
(see in \cite{KauffmanLomonaco2008} for the proof) for the image of the 
homomorphism $\phi$
then the relation of the $B_{3}$ is fulfilled, or we have a representation
of $B_{3}$ into $SU(2)$.

{For more practical scenarios this general 
representation is the following construction%please ensure original meaning is retained
. Let us choose
\[
g=e^{i\theta}=a+b\text{i}
\]
where $a=\cos(\theta)$ and $b=\sin(\theta).$ Let
\[
h=a+b\left[(c^{2}-s^{2}){i}+2cs\cdot{k}\right]
\]
where $c^{2}+s^{2}=1$ and $c^{2}-s^{2}=\frac{a^{2}-b^{2}}{2b^{2}}$.
Then we are able to rewrite $g,h$ as matrices $G,H$. 
In principle, the matrices $G,H$ can be obtained from the expressions
above by a switch of the basis, i.e.,
\[
G=\exp(\theta\boldsymbol{i})\qquad H=a\boldsymbol{1}+b\left[(c^{2}-s^{2})\boldsymbol{i}+2cs\boldsymbol{k}\right]
\]

Here, we choose $H=FGF^{+}$ and 
\[
G=\left(\begin{array}{cc}
e^{i\theta} & 0\\
0 & e^{-i\theta}
\end{array}\right)\qquad F=\left(\begin{array}{cc}
ic & is\\
is & -ic
\end{array}\right)
\]

Among this class of representations, there is the simplest example
\[
g=e^{7\pi i/10},\:f=i\tau+k\sqrt{\tau},\:h=fgf^{-1}
\]
where $\tau^{2}+\tau=1$.} Then, $g,h$ satisfy $ghg=hgh$ the relation
of $B_{3}$. This representation is known as the Fibonacci representation
of $B_{3}$ to $SU(2)$. {The Fibonacci representation is dense in
$SU(2)$, see \cite{topoQuantComp}. This representation is generated by the 20th 
root of unity. Other dense representations are given by $4r$th roots of unity via
recoupling theory, see {\mbox{Sections~1.3 and 1.4}} in \cite{topoQuantComp}.} %MDPI: please confirm if this is the citation of section in reference.

{\section{Knot Group Representations via Berry Phases}
In the previous section we discussed, the representations of the knot group into $SU(2)$ to realize the 1-qubit operations. Central point in this paper is the representation
\[
\phi:\pi_{1}\left(C(K)\right)\to SU(2)
\]
of the knot group. The fundamental group is a topological invariant and 
we have to realize this group in a quantum system. In this section, we will realize these 1-qubit gates, the 2-qubit gate will be described in the next section.
Here, we will discuss the direct realization of the knot group, i.e., the fundamental group of the knot complement. We will not discuss the abstract representation of the group $\pi_1(C(K))$ by quantum gates which is also possible.
For that purpose we have to define the fundamental group more carefully. 
Let $X$ be a topological space or a manifold. A map $\gamma:[0,1]\to X$ with $\gamma(0)=\gamma(1)$ is a closed curve. Two curves $\gamma_1,\gamma_2$ are homotopic $\gamma_1\simeq\gamma_2$ if there is a one-parameter family of continuous maps which deform $\gamma_1$ to $\gamma_2$. The concatenation $\gamma_1\#\gamma_2$ of curves (up to homotopy) is the group operation making the set of homotopy classes of closed curves to a group, the fundamental group $\pi_1(X)$.
Now, we will discuss the representation of the fundamental group by the holonomy along a closed curve, i.e., by an integral of a gauge connection or potential along a closed curve.
As shown by Milnor \cite{MilnorZeroCurvature1958}, there is one-to-one relation between a homomorphism $\pi_1(X)\to G$ into the Lie group $G$ and the integral
\[
\ointop_{\gamma} A \qquad\mbox{with}\qquad dA+A\wedge A=0
\]  
of a flat $G$-connection $A$, i.e., this integral depends only on the homotopy class of the closed curve $\gamma$.
}
In our case, we will interpret the representation
\[
\phi:\pi_{1}\left(C(K)\right)\to SU(2)
\]
up to conjugation as a flat connection of a $SU(2)$ principal bundle
over the knot complement $C(K)$. Let $P$ be a $SU(2)$ principal
fiber bundle over $C(K)$ with connection $A$ locally represented
by a 1-form with values in the adjoint representation of the Lie algebra
$\mathfrak{su}(2)$, i.e., $A\in\Lambda^{1}(C(K))\otimes ad(\mathfrak{su}(2))$.
The connection is flat if the curvature
\[
F=dA+A\wedge A=0
\]
vanishes. In this case (see Milnor \cite{MilnorZeroCurvature1958})
the integral
\[
\ointop_{\gamma}A
\]
along a closed curve $\gamma:S^{1}\to C(K)$ depends only on the homotopy
class $[\gamma]\in\pi_{1}\left(C(K)\right)$ and {the exponential
\[
\pi_1(C(K))\ni\gamma\to\phi(\gamma)=\mathcal{P}\exp\left(\ointop_{\gamma}A\right)\in SU(2)
\]
for varying closed curves ($\mathcal{P}$ path ordering operator)
gives a representation $\pi_{1}\left(C(K)\right)\to SU(2)$.} However, which
quantum system realized this representation? Let us consider the Hamiltonian
\[
H=H_{0}+h
\]
with a non-adiabatic and adiabatic part. The whole Hamiltonian has
to fulfill the usual Schr{\"o}dinger equation
\[
i\hbar\frac{\partial}{\partial t}|\psi\rangle=H|\psi\rangle
\]
and we have to demand that each eigenstate $|k_{n}\rangle$ of the
Hamiltonian $h$ with discrete spectrum
\[
h|k_{n}\rangle=E_{n}|k_{n}\rangle
\]
develops independently in time. Then, we have the decomposition
\[
|\psi\rangle=\sum_{n}a_{n}|k_{n}\rangle
\]
leading to the solution
\[
a_{n}=\exp\left(-\frac{i}{\hbar}\int E_{n}(\tau)d\tau\right)\exp\left(\intop\langle k_{n}|\frac{\partial}{\partial\tau}|k_{n}\rangle d\tau\right)
\]
where the second expression is known as geometric phase or Berry phase.
Usually the states are parameterized by some manifold $M$ with coordinates
$x=x(t)$ and we consider a cyclic evolution $x(0)=x(T)$, i.e., closed
curves in $M$. Then, the Berry phase is given by
\[
\theta_{top}=\ointop_{\gamma}\langle k_{n}|d|k_{n}\rangle
\]
and the expression $A_{Berry}=\langle k_{n}|d|k_{n}\rangle$ as Berry
connection. 
{This solution is well known and for completeness we described it here again. At first, the Berry connection is the connection of $U(1)$ principal bundle. Second, the curvature $\Omega=dA_{Berry}$ is non-zero.
Therefore, at the first view we cannot use this connection to represent the
knot group. We need a connection with values in $SU(2)$ to
get a representation for $\phi$ via $\exp(\theta_{top})$.
Our idea is now to rearrange the components of the Berry connection (including the off-diagonal terms) to produce a $SU(2)$ connection. For that purpose,
we will restrict the system to a 2-level system, $|k_{0}\rangle,|k_{1}\rangle$.
Then, we remark that the Lie algebra $\mathfrak{su}(2)$ is generated
by the three Pauli matrices $\sigma_{x},\sigma_{y},\sigma_{z}$ so
that every element is given by a linear combination $a\cdot\sigma_{x}+b\cdot\sigma_{y}+c\cdot\sigma_{z}$.
Now we arrange the possible connection components $\omega_{nm}=\langle k_{n}|d|k_{m}\rangle$
into one matrix
\[
\omega=\left(\begin{array}{cc}
\langle k_{0}|d|k_{0}\rangle & \langle k_{0}|d|k_{1}\rangle\\
\langle k_{1}|d|k_{0}\rangle & \langle k_{1}|d|k_{1}\rangle
\end{array}\right)
\]
with the decomposition
\[
\omega=\langle k_{0}|d|k_{0}\rangle\frac{\boldsymbol{1}+\sigma_{z}}{2}+\langle k_{1}|d|k_{1}\rangle\frac{\boldsymbol{1}-\sigma_{z}}{2}+\Re(\langle k_{0}|d|k_{1}\rangle)\sigma_{x}+\Im\langle k_{0}|d|k_{1}\rangle\sigma_{y}
\]

By using the normalization $\langle k_{n}|k_{m}\rangle=\delta_{nm}$
one gets
\[
0=d\left(\langle k_{n}|k_{m}\rangle\right)=\omega_{nm}^{*}+\omega_{nm}
\]
and
\[
d\omega_{nm}=\sum_{k}\omega_{nk}^{*}\wedge\omega_{km}=-\sum_{k}\omega_{nk}\wedge\omega_{km}
\]
so that we obtain for the curvature $\Omega$
\[
\Omega=d\omega+\omega\wedge\omega=0
\]

Obviously $\omega$ is a connection of a flat $SU(2)$ bundle and
the integral
\[
\ointop_{\gamma}\omega
\]
depends only on the homotopy class of the closed curve $\gamma$, as we want.
The off-diagonal terms like $\langle k_{0}|d|k_{1}\rangle$ of the connection $\omega$ can be calculated with respect to the expectation values of $dh$. 
Together with the eigenvalues $E_0,E_1$ of $h$ for $|k_0\rangle,|k_1\rangle$, respectively, we obtain, for instance,
\[
\langle k_{0}|d|k_{1}\rangle=\frac{\langle k_{0}|dh|k_{1}\rangle}{E_0-E_1}
\]}

Then, the exponential of this integral gives a representation $\phi$
of the fundamental group into $SU(2).$ 
{Now we go back to the trefoil knot complement $C(3_1)$ 
with fundamental group $\pi_1(C(3_1))=B_{3}$. Then, the Berry phases
along the two generators $a,b$ (i.e., two closed curves) of the fundamental
group $\pi_{1}\left(C(3_{1})\right)=B_{3}$ generate the 1-Qubit operations.
Via the Berry phases, these operations act on the quantum system to
influence its state. Keeping this idea in mind, we have the following scheme: 
consider a qubit on the trefoil knot and consider the two generators $a,b$ of the knot group (see Figure \ref{fig:generator-trefoil}). 

If we do manipulations along these two closed curves we are able to influence the qubit by using the Berry phase. Here, we refer to the work \cite{BerryPhaseQC2014} for ideas to use the Berry phase for quantum computing.
} 

\section{Linking and 2-Qubit Operations\label{sec:Linking-and-2-qubit}}

In the previous section, we described the appearance of the braid
group $B_{3}$ as fundamental group of the trefoil knot complement.
{Above we described the situation that the knot complement 
of the trefoil knot determines the operations or quantum gates. 
In this case, the quantum
gates are braiding operations (used for anyons) for three-strand braids.
Unfortunately, the $SU(2)-$representations of the braid group $B_{3}$
are only 1-qubit gates but one needs at least a 2-qubit gate like
CNOT to represent any quantum circuit.}
It is known that for 2-qubit operations, one needs elements of the
braid group $B_{6}$. Is there other knot complements having braid
groups as fundamental groups? Unfortunately, the answer is no. Here,
is the line of arguments: every knot complement is determined by the
fundamental group (aspherical space), then the cohomology of knot
complements is determined by the first two groups ($0$th and $1$th),
all other groups are given by duality. But the braid groups $B_{n}$
for $n>3$ have non-trivial cohomology groups in degree 3 or higher
which is impossible for knot complements.

\begin{figure}
%\widefigure
\includegraphics[scale=0.7]{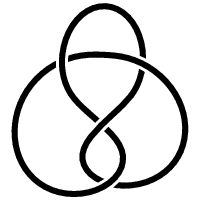}\caption{figure-8 knot $4_{1}$. \label{fig:figure-8-knot}}
\end{figure}

\begin{figure}
%\widefigure
\includegraphics[scale=0.4]{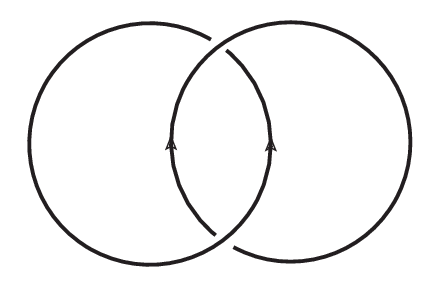}

\caption{Hopf link $L2a1$.\label{fig:Hopf-link}}

\end{figure}

In the previous sections, we describe the knot complement of the simplest
knot, the trefoil. However, there are more complicated knots. The complexity
of knots is measured by the number of crossings. There is only one
knot with three crossings (trefoil) and with four crossings (figure-8).
For the figure-8 knot $4_{1}$ (see Figure \ref{fig:figure-8-knot}),
the knot group is given by
\[
\pi_{1}\left(C(4_{1})\right)=\langle a,b\,|\:bab^{-1}ab=aba^{-1}ba\rangle
\]
admitting a representation $\phi$ into $SU(2)$, see \cite{KirKla:90}. 
{Here, we remark that the figure-8 knot is part of a large class, the so-called hyperbolic knots. Hyperbolic knots are characterized by the property that the knot complement admits a hyperbolic geometry. Hyperbolic knot complements have special properties, in particular topology and geometry are connected in a special way. We will come back to these ideas in our forthcoming work.
As explained above, knot groups admit representations into $SU(2)$ leading to 1-qubit operations. }
Therefore,  we have to change the complexity in another direction by
adding more components, i.e., we have to go from knots to links. The simplest link
is the Hopf link (denoted as $L2a1$, see Figure \ref{fig:Hopf-link}),
the linking of two unknotted curves. 
The knot group is simply
\[
\pi_{1}\left(C(L2a1)\right)=\langle a,b\,|\,aba^{-1}b^{-1}=[a,b]=e\rangle=\mathbb{Z}\oplus\mathbb{Z}
\]

\begin{figure}
%\widefigure
\includegraphics[scale=0.25]{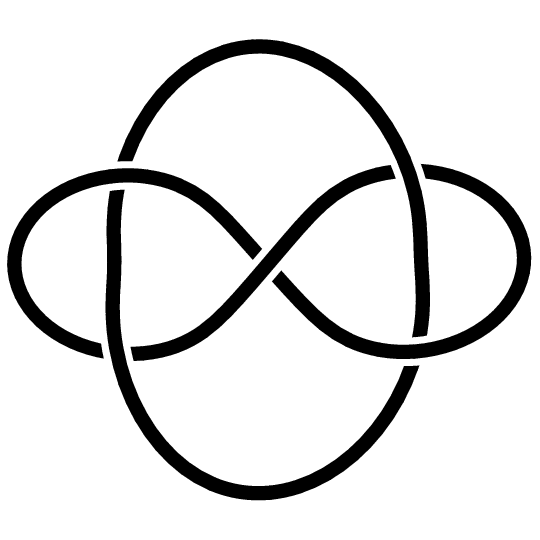}
\caption{Whitehead link $Wh$. \label{fig:Whitehead-link} }
\end{figure}

Here, we will discuss a toy model, every component is related to a
$SU(2)$ representation, i.e., the knot group $\pi_{1}(C(L2a1))$ is
represented as $SU(2)\otimes SU(2)$ via the Berry connection. Now
we associate to each component of the link a representation and a
generator of $SU(2)$ (i.e., $\sigma_{x},\sigma_{y}$ or $\sigma_{z}$),
say $\sigma_{x}$ to one component (the generator $a\in\pi_{1}\left(C(L2a1)\right)$)
and $\sigma_{z}$ to the other component (the generator $b\in\pi_{1}\left(C(L2a1)\right)$).
{By using the relation between the group commutator 
and the Lie algebra commutator of the enveloped Lie algebra $U(SU(2))$, 
we want to express the relation $[a,b]=e$ via the exponential of the commutator 
$\sigma_{x}\otimes\sigma_{z}-\sigma_{z}\otimes\sigma_{x}$ via the usual relation 
between the Lie algebra commutator and this commutator. It induces a
representation
$$
\pi_{1}\left(C(L2a1)\right)\to SU(2)\otimes SU(2)
$$
by using the exponential map $\exp (\mathfrak{su}(2)\otimes\mathfrak{su}(2))$.
The relation in $\pi_{1}\left(C(L2a1)\right)$ can be expressed as 
Lagrangian multiplier (in the usual way) 
so that we get the Hamiltonian} 
\[
H=\sigma_{x}\otimes\sigma_{z}-\sigma_{z}\otimes\sigma_{x}
\]
and we get the qubit operation by the exponential
\[
U=\exp(it(\sigma_{x}\otimes\sigma_{z}-\sigma_{z}\otimes\sigma_{x}))
\]
for a suitable time $t$. Now we see the principle: we associate a
term $\sigma_{x}\otimes\sigma_{z}$ to an over-crossing between the
two components and a term $\sigma_{z}\otimes\sigma_{x}$ for the under-crossing.
In the last example, we will consider the famous Whitehead link (see Figure
\ref{fig:Whitehead-link}). 
The knot group is given by

\[
\pi_{1}\left(C(Wh)\right)=\langle x,y\:|\:[x,y][x,y^{-1}][x^{-1},y^{-1}][x^{-1},y]=e\rangle
\]
and as described above we will associate the tensor products of the
generators to the over-crossings or under-crossings between the components.
Then, we will get 
\[
H=2\sigma_{x}\otimes\sigma_{z}-2\sigma_{z}\otimes\sigma_{x}
\]
with the operation
\[
U=\exp(i2t(\sigma_{x}\otimes\sigma_{z}-\sigma_{z}\otimes\sigma_{x}))
\]
with another choice of the time variable.

\section{Discussion}

In this paper, we presented some ideas to use 3-manifolds for quantum
computing. A direct usage for surfaces (related to anyons) is not
possible, but we explained above that the best representative is the
fundamental group of a manifold. The fundamental group is the set
of closed curves up to deformation with concatenation as group operation
(also up to deformation). Every 3-manifold can be decomposed into
simple pieces so that every piece carries a geometric structure (out
of eight classes). In principle, the pieces consist of complements of
knots and links. Then, the fundamental group of the knot complement,
known as knot group, is an important invariant of the knot or link.
Why not use this knot group for quantum computing? In \cite{Planat2017,Planat2017a,Planat2018,Planat2019},
M. Planat et al. studied the representation of knot groups and the
usage for quantum computing. Here, we discussed a direct relation between
the knot complement and quantum computing via the Berry phase. The
knot group determines the operations where we fix a suitable $SU(2)$
representation. As a result, we get all 1-qubit operations for a knot.
Then, we discussed the construction of 2-qubit operations by the linking
of two knots. The concrete realization of these ideas by a device
will be shifted to our forthcoming work. 
%\end{paracol}

\section*{Acknowledgments}
{I acknowledge useful discussions with Michel Planat. Furthermore, I acknowledge the helpful remarks and questions of the referees leading to better readability of this paper.}
%\conflictsofinterest{The author declares no conflicts of interest.}

%\end{paracol}
%\reftitle{References}
%\bibliographystyle{Definitions/mdpi}
%\bibliography{../foliation-gerbes,../diffbib,../knots,../quantum-theory,../measurements,../quantum,../exbib-for-carl,../exbib_new}

\end{document}